\documentclass[letter]{IEEEtran}
\usepackage[T1]{fontenc}
\usepackage[latin9]{inputenc}
\usepackage{verbatim}
\usepackage{float}
\usepackage{amsmath}
\usepackage{amsthm}
\usepackage{amssymb}
\usepackage{etoolbox}

\theoremstyle{plain}
\newtheorem{thm}{\protect\theoremname}

\usepackage{cite}

\usepackage{epstopdf}
\usepackage{xfrac}
\usepackage[ruled]{algorithm2e}

\DeclareMathOperator{\tr}{tr}

\DeclareMathOperator{\diag}{diag}

\newcommand{\herm}{^{{\dagger}}}
\newcommand{\trans}{^{\mathsf{T}}}
\DeclareMathOperator{\argmin}{argmin}

\usepackage{subfig}


\usepackage{babel}
\providecommand{\theoremname}{Theorem}

\begin{document}
\IEEEoverridecommandlockouts
\title{\huge On the Optimality of the Stationary Solution of Secrecy Rate Maximization for MIMO Wiretap
Channel}
\author{Anshu Mukherjee,~\IEEEmembership{Student Member, IEEE}, Vaibhav
Kumar, \IEEEmembership{Member, IEEE}, Eduard Jorswieck,~\IEEEmembership{Fellow, IEEE}, Björn Ottersten,~\IEEEmembership{Fellow, IEEE},
and Le-Nam Tran,~\IEEEmembership{Senior Member, IEEE}\thanks{A. Mukherjee, V. Kumar, and L.-N. Tran are with School of Electrical
and Electronic Engineering, University College Dublin, Ireland. Email:
anshu.mukherjee@ucdconnect.ie; vaibhav.kumar@ieee.org; nam.tran@ucd.ie.}
\thanks{E. Jorswieck is with  the Institute of Communications Technology,
	Technische Universität Braunschweig, Germany. Email: Jorswieck@ifn.ing.tu-bs.de.}
\thanks{B. Ottersten is with Interdisciplinary Centre for Security, Reliability
and Trust, University of Luxembourg, Luxembourg. Email: bjorn.ottersten@uni.lu.}%
%\thanks{This publication has emanated from research supported in part by a
%Grant from Science Foundation Ireland under Grant number 17/CDA/4786.}%
}
%\IEEEaftertitletext{\vspace{-2\baselineskip}}
\maketitle
\begin{abstract}
%This paper considers the secrecy capacity of a multiple-input multiple-output
%(MIMO) Gaussian wiretap channel (WTC), which involves solving a difference
%of convex functions program known to be non-convex for the non-degraded
%case.
To achieve perfect secrecy in a multiple-input multiple-output
(MIMO) Gaussian wiretap channel (WTC), we need to find its secrecy capacity and  optimal signaling, which involves solving a difference
of convex functions program known to be non-convex for the non-degraded
case.
 To deal with this, a class of existing solutions have been developed
but only local optimality is guaranteed by standard convergence analysis.
Interestingly, our extensive numerical experiments have shown that
these local optimization methods \emph{indeed achieve global optimality}.
In this paper, we provide an analytical proof for this observation.
To achieve this, we show that the Karush-Kuhn-Tucker (KKT) conditions
of the secrecy rate maximization problem admit a unique solution for \emph{both degraded and non-degraded cases}.
Motivated by this, we also propose a low-complexity algorithm to find
a stationary point. Numerical results are presented to verify the
theoretical analysis.
\end{abstract}


\section{Introduction}

\allowdisplaybreaks

\IEEEPARstart{W}{ith} the advent of new wireless communication applications
including social networking, financial transactions, and military-related
communications, there has been an ever increasing demand for privacy-preserving
communication services. Traditional data security schemes such as
cryptography are implemented at a higher layer of the communication
network. However, Wyner in his seminal work introduced an information-theoretic
paradigm of physical layer security (PLS) for discrete memoryless
wiretap channel (WTC)~\cite{Wyner75}. Building on~\cite{Wyner75},
the authors in~\cite{Gauss_wiretap} derived the secrecy capacity
of a Gaussian WTC, and established that the achievable rate region
of a Gaussian WTC can be completely characterized by the corresponding
secrecy capacity.

It was established in~\cite{ZangLi} that the secrecy rate maximization
(SRM) problem for the general multiple-input multiple-output (MIMO)
Gaussian WTC is non-convex and thus is difficult to solve. The authors
in~\cite{ZangLi} therefore considered a special case of multiple-input
single-output (MISO) Gaussian WTC with perfect instantaneous channel
state information at the transmitter (CSIT), and derived an analytical
solution for the (reformulated) optimization problem. In~\cite{Secrecy_cap_MISOME},
a beamforming strategy was applied and was shown to be secrecy capacity-achieving
for the case of a multi-antenna transmitter, a single-antenna receiver
and a  multi-antenna eavesdropper (MISOME) system. The problem of computing
the secrecy capacity for the case of a multi-antenna transmitter, a multi-antenna
receiver and a multi-antenna eavesdropper (MIMOME) was studied in~\cite{MIMOME_WTC},
where the authors showed that  Gaussian signaling indeed achieves
the secrecy capacity and also derived the optimal covariance structure.
Independently, the secrecy capacity of the MIMO WTC was also studied
in~\cite{Oggier2011SecCapEq}. There is indeed a rich literature
concerning analytical and numerical solutions to find the secrecy
capacity of the MIMO WTC channel \cite{Sec_MIMO_SPC_3,Li2013b,MIMO_POTDC,lyoka2015minmaxEQ,Loyka2016,ThangNguyen2020,Anshu:MIMOWTC:21VTC,Anshu:MIMOWTC:2020}.%
\begin{comment}
We can reduce the discussion on existing literature since for SPL,
we only have 4 pages for technical discussions. We can include as
many references as we want on the last page.
\end{comment}

It is well known that the secrecy capacity problem for the non-degraded
MIMO WTC is non-convex in the original form.\footnote{The MIMO WTC is said to be non-degraded if the difference between the receiver's channel and the eavesdropper's channel is indefinite.} Consequently, there has
been a concern that algorithms based on convex optimization techniques
applied to the original form can be trapped in a locally optimal solution
far from the secrecy capacity. To avoid this, the equivalent convex-concave
reformulation of the secrecy capacity problem has been used to find
the optimal signaling for the non-degraded MIMO WTC \cite{Oggier2011SecCapEq,lyoka2015minmaxEQ}.
Against this background, the main contributions in this letter are
as follows:
\begin{itemize}
\item We give a rigorous analytical proof that there exists \emph{a unique
Karush--Kuhn--Tucker (KKT) point} for  the secrecy
capacity problem for  both degraded and non-degraded Gaussian MIMO WTC.
 This interesting result in fact establishes
that existing local optimization methods such as \cite{Li2013b} for
the secrecy problem indeed yield the globally-optimal solution.
\item Motivated by this result, we propose an accelerated gradient projection algorithm
with adaptive momentum parameters that solves the secrecy capacity
problem directly, rather than the equivalent convex-concave form.
The proposed algorithm is provably convergent to a KKT point, and thus solves the MIMOME precoding problem globally and efficiently.  
\end{itemize}
\textit{Notation:} We use bold uppercase and lowercase letters to
denote matrices and vectors, respectively. $\tr\left(\mathbf{X}\right),\left|\mathbf{X}\right|$
and $\left\Vert \mathbf{X}\right\Vert $ denote the trace, determinant
and Frobenius norm of $\mathbf{X}$, respectively. By $\mathbf{X}_{i,j}$
we denote the $j$-th element of the $i$-th row of matrix $\mathbf{X}$.
$\mathbf{\left(\cdot\right)}\herm$ and $\mathbf{\left(\cdot\right)\trans}$
denote the Hermitian transpose and (ordinary) transpose, respectively.
$\mathbb{C}^{M\times N}$ denotes the space of complex matrices of
size $M\times N$, and $\mathbb{E}\left\{ \cdot\right\} $ is the
expectation operator. $\diag\left(\mathbf{x}\right)$ denotes the
square diagonal matrix which has the elements of $\mathbf{x}$ on
the main diagonal, and $\left[x\right]_{+}\triangleq\max\left\{ x,0\right\} $.
$\mathbf{I}$ and $\mathbf{0}$ represent identity and zero matrices,
respectively. By $\mathbf{A}\succeq(\succ)\mathbf{B}$ we mean that
$\mathbf{A}-\mathbf{B}$ is positive semidefinite (definite). The
maximum eigenvalue of $\mathbf{X}$ is denoted by $\sigma_{\max}(\mathbf{X})$.

\section{System Model and Secrecy Capacity Problem}

We consider a communication system that includes Alice as the transmitter,
Bob as the legitimate receiver, and Eve as the eavesdropper. In this MIMO
system, Alice wants to transmit  information to Bob in the
presence of Eve, where Alice is equipped with $N_{t}$ number of transmitting
antenna, Bob and Eve are equipped with $N_{r}$ and $N_{e}$ number
of antennas respectively. Let us denote, $\mathbf{H}\in\mathbb{C}^{N_{r}\times N_{t}}$
and $\mathbf{G}\in\mathbb{C}^{N_{e}\times N_{t}}$ as the channel
matrices corresponding to Bob and Eve. The received signal at legitimate
receiver and at eavesdropper can be expressed as
\begin{equation}
\mathbf{y}_{b}=\mathbf{H}\mathbf{q}+\mathbf{z}_{b};\ \mathbf{y}_{e}=\mathbf{G}\mathbf{q}+\mathbf{z}_{e}
\end{equation}
where $\mathbf{q}$ is the transmitted signal $\mathbf{q}\in\mathbb{C}^{N_{t}\times1}$.
The additive white Gaussian noise at the legitimate receiver and at
the eavesdropper represented as $\mathbf{z}_{b}\in\mathbb{C}^{N_{r}\times1}\sim\mathcal{CN}(\mathbf{0,I})$
and $\mathbf{z}_{e}\in\mathbb{C}^{N_{e}\times1}\sim\mathcal{CN}(\mathbf{0,I})$.
\foreignlanguage{american}{I}n this paper, we assume that $\mathbf{H}$
and $\mathbf{G}$ are 
%to be quasi-static and 
perfectly known at Alice
and Bob. Let $\mathbf{Q}=\mathbb{E}\{\mathbf{q}\mathbf{q}\herm\}\succeq\mathbf{0}$
be the input covariance matrix. Then the secrecy capacity under a
sum power constraint (SPC) has been expressed as \cite{Oggier2011SecCapEq}
\begin{equation}
C_{s}=\max\ \bigl\{ C_{s}(\mathbf{Q})\ \bigl|\ \mathbf{Q}\in\mathcal{Q}\bigr\}\label{eq:secrecycapacity:org}
\end{equation}
where $\mathcal{Q}=\bigl\{\mathbf{Q}\ \bigl|\ \tr(\mathbf{Q})\leq P_{T};\mathbf{Q}\succeq\mathbf{0}\bigr\}$,
$C_{s}(\mathbf{Q})=\ln|\mathbf{I}+\mathbf{H}\mathbf{Q}\mathbf{H}\herm|-\ln|\mathbf{I}+\mathbf{G}\mathbf{Q}\mathbf{G}\herm|$,
$P_{T}>0$ is the total transmit power. Problem (\ref{eq:secrecycapacity:org})
is non-convex in general.

\section{Uniqueness of KKT point of (\ref{eq:secrecycapacity:org})}

We remark that if $\mathbf{H}\herm\mathbf{H}-\mathbf{G}\herm\mathbf{G}$ is negative semi-definite, $ C_{s} $ is zero. Thus, the next theorem provides a complete characterization of the uniqueness of the stationary point of~(\ref{eq:secrecycapacity:org}).
\begin{thm}
\label{thm:uniqueKKT}Assume $\mathbf{H}\herm\mathbf{H}-\mathbf{G}\herm\mathbf{G}$
is  not negative semi-definite. Then problem (\ref{eq:secrecycapacity:org})
has a unique KKT point. Therefore the KKT conditions are necessary and sufficient for the optimality of problem (\ref{eq:secrecycapacity:org}). 
\end{thm}
\begin{IEEEproof}
The Lagrangian function of problem (\ref{eq:secrecycapacity:org})
is 
\begin{equation}
\mathcal{L}(\mathbf{Q},\mathbf{Z},\lambda)=C_{s}(\mathbf{Q})-\lambda\bigl(\tr(\mathbf{Q})-P_{T}\bigr)+\tr(\mathbf{Q}\mathbf{Z})
\end{equation}
where $\lambda\geq0$ and $\mathbf{Z}\succeq\mathbf{0}$ are the Lagrangian
multiplier for the constraints $\tr(\mathbf{Q})\leq P_{T}$ and $\mathbf{Q}\succeq\mathbf{0}$,
respectively. Note that the gradient of $C_{s}(\mathbf{Q})$ is
\begin{equation}
\nabla C_{s}(\mathbf{Q})=\mathbf{H}\herm\bigl(\mathbf{I}+\mathbf{H}\mathbf{\mathbf{Q}}\mathbf{H}\herm\bigr)^{-1}\mathbf{H}-\mathbf{G}\herm\bigl(\mathbf{I}+\mathbf{G}\mathbf{\mathbf{Q}}\mathbf{G}\herm\bigr)^{-1}\mathbf{G}.\label{eq:gradF}
\end{equation}
Thus, the KKT conditions of (\ref{eq:secrecycapacity:org}) are given
by\begin{IEEEeqnarray}{c}
\mathbf{H}\herm\bigl(\mathbf{I}\!+\!\mathbf{H}\mathbf{Q}\mathbf{H}\herm\bigr)^{-1}\mathbf{H}\!-\!\mathbf{G}\herm\bigl(\mathbf{I}\!+\!\mathbf{G}\mathbf{Q}\mathbf{G}\herm\bigr)^{-1}\mathbf{G}\!-\!\lambda\mathbf{I}\!+\!\mathbf{Z}\!=\!\mathbf{0}\IEEEeqnarraynumspace \IEEEyesnumber \IEEEyessubnumber 
\label{eq:KKT1}\\
\tr(\mathbf{Q})\leq P_{T};\lambda\geq0;\lambda\bigl(\tr(\mathbf{Q})-P_{T}\bigr)=0\IEEEyessubnumber \label{eq:KKT2}\\ \mathbf{Z}\succeq\mathbf{0};\mathbf{Q}\succeq\mathbf{0};\mathbf{Q}\mathbf{Z}=\mathbf{0}.\IEEEyessubnumber \label{eq:KKT5}
\end{IEEEeqnarray}Throughout the proof we use the following equality which
is a special case of the matrix inversion lemma \cite[Eqn. (2.8.20)]{Bernstein:2009}
\begin{equation}
\mathbf{Y}(\mathbf{I}+\mathbf{X}\mathbf{Y})^{-1}=(\mathbf{I}+\mathbf{Y}\mathbf{X})^{-1}\mathbf{Y}.\label{eq:matrixinverse}
\end{equation}
In the first part of the proof we assert that $\lambda>0$ and thus
$\tr(\mathbf{Q})=P_{T}$ if $\mathbf{Q}$ is a KKT point. To proceed
we note that applying (\ref{eq:matrixinverse}) to (\ref{eq:KKT1})
yields 
\begin{equation}
(\mathbf{I}+\mathbf{H}\herm\mathbf{H}\mathbf{Q})^{-1}\mathbf{H}\herm\mathbf{H-\mathbf{G}\herm\mathbf{G}(\mathbf{I}+\mathbf{Q}\mathbf{G}\herm\mathbf{G})^{-1}=\lambda\mathbf{I}-\mathbf{Z}.}\label{eq:KKT11}
\end{equation}
Suppose to the contrary that $\lambda=0$. Then (\ref{eq:KKT11})
reduces to 
\begin{equation}
(\mathbf{I}+\mathbf{H}\herm\mathbf{H}\mathbf{Q})^{-1}\mathbf{H}\herm\mathbf{H}-\mathbf{G}\herm\mathbf{G}(\mathbf{I}+\bar{\mathbf{Q}}\mathbf{G}\herm\mathbf{G})^{-1}=-\mathbf{Z}
\end{equation}
which is equivalent to
\begin{equation}
\mathbf{H}\herm\mathbf{H}-\mathbf{G}\herm\mathbf{G}=-\bigl(\mathbf{I}+\mathbf{H}\herm\mathbf{H}\mathbf{Q}\bigr)\mathbf{Z}\bigl(\mathbf{I}+\mathbf{Q}\mathbf{G}\herm\mathbf{G}\bigr).\label{eq:KKT:stationary:2}
\end{equation}
It is clear that the right hand side of (\ref{eq:KKT:stationary:2})
is negative semidefinite, which contradicts the assumption that $\mathbf{H}\herm\mathbf{H}-\mathbf{G}\herm\mathbf{G}$
is positive semidefinite or indefinite. Thus we can conclude that
$\lambda>0$ and thus $\tr(\mathbf{Q})=P_{T}$.

Next, we show that there is a unique solution $\mathbf{Q}$ to the
KKT conditions. Suppose to the contrary that $(\mathbf{Q}_{1},\mathbf{Z}_{1},\lambda_{1})$
and $(\mathbf{Q}_{2},\mathbf{Z}_{2},\lambda_{2})$ are two different
KKT points of (\ref{eq:secrecycapacity:org}). Let us define 
\begin{IEEEeqnarray}{c}
\boldsymbol{\Phi}_{i}=\mathbf{H}\herm\bigl(\mathbf{I}+\mathbf{H}\mathbf{Q}_{i}\mathbf{H}\herm\bigr)^{-1}\mathbf{H};\boldsymbol{\Psi}_{i}=\mathbf{G}\herm\bigl(\mathbf{I}+\mathbf{G}\mathbf{Q}_{i}\mathbf{G}\herm\bigr)^{-1}\mathbf{G}\IEEEeqnarraynumspace
\end{IEEEeqnarray}
for $i=\{1,2\}$. Then (\ref{eq:KKT1}) for those two KKT points is
\begin{subequations}\label{eq:KKT:stationary:reduced}
\begin{align}
\mathbf{Z}_{1}+\boldsymbol{\Phi}_{1}-\boldsymbol{\Psi}_{1}= & \lambda_{1}\mathbf{I}\label{eq:KKT:stationary:reduced1}\\
\mathbf{Z}_{2}+\boldsymbol{\Phi}_{2}-\boldsymbol{\Psi}_{2}= & \lambda_{2}\mathbf{I}.\label{eq:KKT:stationary:reduced2}
\end{align}
\end{subequations} Since, $\tr(\mathbf{Q}_{1})=\tr(\mathbf{Q}_{2})=P_{T}$,
it is easy to check that \begin{IEEEeqnarray}{rcl}
\tr\left(\!\bigl(\boldsymbol{\Phi}_{1}-\boldsymbol{\Psi}_{1}+\mathbf{Z}_{1}\bigr)\bigl(\mathbf{Q}_{1}-\mathbf{Q}_{2}\bigr)\!\right)&=&\lambda_{1}\tr\bigl(\mathbf{Q}_{1}-\mathbf{Q}_{2}\bigr)=0,\label{eq:Stp1_1} 
\IEEEeqnarraynumspace 
\end{IEEEeqnarray}\begin{IEEEeqnarray}{rcl}
\tr\left(\!\bigl(\boldsymbol{\Phi}_{2}-\boldsymbol{\Psi}_{2}+\mathbf{Z}_{2}\bigr)\bigl(\mathbf{Q}_{1}-\mathbf{Q}_{2}\bigr)\!\right) &=&\lambda_{2}\tr\bigl(\mathbf{Q}_{1}-\mathbf{Q}_{2}\bigr)=0.\label{eq:Stp1_2}
\IEEEeqnarraynumspace 
\end{IEEEeqnarray} Now, subtracting (\ref{eq:Stp1_2}) from (\ref{eq:Stp1_1}),   we obtain{\small 
\thinmuskip=0mu 
\medmuskip =0mu
\thickmuskip =0mu
\begin{IEEEeqnarray}{c}
\tr\bigl(\bigl(\boldsymbol{\Phi}_{1}-\boldsymbol{\Phi}_{2}-\bigl(\boldsymbol{\Psi}_{1}-\boldsymbol{\Psi}_{2}\bigr)\bigr)\bigl(\mathbf{Q}_{1}-\mathbf{Q}_{2}\bigr)\bigr)-\tr\bigl(\mathbf{Z}_{1}\mathbf{Q}_{2}+\mathbf{Z}_{2}\mathbf{Q}_{1}\bigr)  = 0.\IEEEeqnarraynumspace \label{eq:StarredEq} 
\end{IEEEeqnarray} } Our purpose in the sequel is to show that (\ref{eq:StarredEq})
is impossible if $\mathbf{Q}_{1}\neq\mathbf{Q}_{2}$. To this end
we first express $\boldsymbol{\Phi}_{1}-\boldsymbol{\Phi}_{2}$ as
in (\ref{eq:Phi1minusPhi2}) shown at the top of the following page.
\begin{figure*}[th]
\begin{subequations}\label{eq:Phi1minusPhi2}
\begin{align}
\boldsymbol{\Phi}_{1}-\boldsymbol{\Phi}_{2} & =\mathbf{H}\herm\bigl(\mathbf{I}+\mathbf{H}\mathbf{Q}_{1}\mathbf{H}\herm\bigr)^{-1}\mathbf{H}-\mathbf{H}\herm\bigl(\mathbf{I}+\mathbf{H}\mathbf{Q}_{2}\mathbf{H}\herm\bigr)^{-1}\mathbf{H}=\mathbf{H}\herm\bigl(\bigl(\mathbf{I}+\mathbf{H}\mathbf{Q}_{1}\mathbf{H}\herm\bigr)^{-1}-\bigl(\mathbf{I}+\mathbf{H}\mathbf{Q}_{2}\mathbf{H}\herm\bigr)^{-1}\bigr)\mathbf{H}\label{eq:Stp3_2}\\
 & =\mathbf{H}\herm\bigl(\mathbf{I}+\mathbf{H}\mathbf{Q}_{1}\mathbf{H}\herm\bigr)^{-1}\mathbf{H}\bigl(\mathbf{Q}_{2}-\mathbf{Q}_{1}\bigr)\mathbf{H}\herm\bigl(\mathbf{I}+\mathbf{H}\mathbf{Q}_{2}\mathbf{H}\herm\bigr)^{-1}\mathbf{H}=\boldsymbol{\Phi}_{1}\bigl(\mathbf{Q}_{2}-\mathbf{Q}_{1}\bigr)\boldsymbol{\Phi}_{2}\label{eq:Phi1minusPhi2:3}\\
 & =\bigl(\boldsymbol{\Psi}_{1}+\lambda_{1}\mathbf{I}-\mathbf{Z}_{1}\bigr)\bigl(\mathbf{Q}_{2}-\mathbf{Q}_{1}\bigr)\bigl(\boldsymbol{\Psi}_{2}+\lambda_{2}\mathbf{I}-\mathbf{Z}_{2}\bigr)\label{eq:Phi1minusPhi2:4}
\end{align}
\end{subequations}%
\setcounter{equation}{16}%
 \begin{IEEEeqnarray}{c}{\small
	(\boldsymbol{\Phi}_{1}-\boldsymbol{\Phi}_{2})-(\boldsymbol{\Psi}_{1}-\boldsymbol{\Psi}_{2})=\boldsymbol{\Psi}_{1}\bigl(\mathbf{Q}_{2}-\mathbf{Q}_{1}\bigr)\bigl(\lambda_{2}\mathbf{I}-\mathbf{Z}_{2}\bigr)
	 +\bigl(\lambda_{1}\mathbf{I}-\mathbf{Z}_{1}\bigr)\bigl(\mathbf{Q}_{2}-\mathbf{Q}_{1}\bigr)\boldsymbol{\Psi}_{2}+\bigl(\lambda_{1}\mathbf{I}-\mathbf{Z}_{1}\bigr)\bigl(\mathbf{Q}_{2}-\mathbf{Q}_{1}\bigr)\bigl(\lambda_{2}\mathbf{I}-\mathbf{Z}_{2}\bigr).\IEEEeqnarraynumspace\label{eq:Stp3_3} }
\end{IEEEeqnarray} 
\setcounter{equation}{17}%
\begin{multline}
\tr\left(\left(\boldsymbol{\Phi}_{1}-\boldsymbol{\Phi}_{2}-\bigl(\boldsymbol{\Psi}_{1}-\boldsymbol{\Psi}_{2}\bigr)\right)\bigl(\mathbf{Q}_{1}-\mathbf{Q}_{2}\bigr)\right)=\tr\bigl((\lambda_{1}\mathbf{Q}_{1}+\boldsymbol{\Psi}_{1}\mathbf{Q}_{1})\mathbf{Z}_{2}\mathbf{Q}_{1}\bigr)+\tr\bigl((\lambda_{2}\mathbf{Q}_{2}+\boldsymbol{\Psi}_{2}\mathbf{Q}_{2})\mathbf{Z}_{1}\mathbf{Q}_{2}\bigr)\\
-\lambda_{1}\tr\bigl((\mathbf{Q}_{2}-\mathbf{Q}_{1})\boldsymbol{\Psi}_{2}(\mathbf{Q}_{2}-\mathbf{Q}_{1})\bigr)-\lambda_{2}\tr\bigl((\mathbf{Q}_{2}-\mathbf{Q}_{1})\boldsymbol{\Psi}_{1}(\mathbf{Q}_{2}-\mathbf{Q}_{1})\bigr)-\lambda_{1}\lambda_{2}\tr\bigl((\mathbf{Q}_{2}-\mathbf{Q}_{1})(\mathbf{Q}_{2}-\mathbf{Q}_{1})\bigr)\label{eq:13}
\end{multline}
\setcounter{equation}{22}%
\begin{multline}
\tr\bigl((\lambda_{1}\mathbf{Q}_{1}+\boldsymbol{\Psi}_{1}\mathbf{Q}_{1})\mathbf{Z}_{2}\mathbf{Q}_{1}\bigr)+\tr\bigl((\lambda_{2}\mathbf{Q}_{2}+\boldsymbol{\Psi}_{2}\mathbf{Q}_{2})\mathbf{Z}_{1}\mathbf{Q}_{2}\bigr)-\tr(\mathbf{Z}_{2}\mathbf{Q}_{1}+\mathbf{Z}_{1}\mathbf{Q}_{2})\\-\lambda_{1}\tr\bigl((\mathbf{Q}_{2}-\mathbf{Q}_{1})\boldsymbol{\Psi}_{2}(\mathbf{Q}_{2}-\mathbf{Q}_{1})\bigr)
-\lambda_{2}\tr\bigl((\mathbf{Q}_{2}-\mathbf{Q}_{1})\boldsymbol{\Psi}_{1}(\mathbf{Q}_{2}-\mathbf{Q}_{1})\bigr)-\lambda_{1}\lambda_{2}\tr\bigl((\mathbf{Q}_{2}-\mathbf{Q}_{1})(\mathbf{Q}_{2}-\mathbf{Q}_{1})\bigr)\leq0\label{eq:14}
\end{multline}
\vspace{-5pt}\hrulefill
\end{figure*}
 Note that in (\ref{eq:Phi1minusPhi2:3}) we have used the fact that
$\mathbf{X}^{-1}-\mathbf{Y}^{-1}=\mathbf{X}^{-1}\bigl(\mathbf{Y}-\mathbf{X}\bigr)\mathbf{Y}^{-1}$
for invertible matrices $\mathbf{X}$ and $\mathbf{Y}$, and that
(\ref{eq:Phi1minusPhi2:4}) is due to (\ref{eq:KKT:stationary:reduced}).
Similarly we can write \setcounter{equation}{15}
\begin{equation}
\boldsymbol{\Psi}_{1}-\boldsymbol{\Psi}_{2}=\boldsymbol{\Psi}_{1}\bigl(\mathbf{Q}_{2}-\mathbf{Q}_{1}\bigr)\boldsymbol{\Psi}_{2}.\label{eq:Stp3_1}
\end{equation}
Combining (\ref{eq:Stp3_1}) with (\ref{eq:Phi1minusPhi2:4}) produces \eqref{eq:Stp3_3} shown at the top of the following page.
%{\small \begin{IEEEeqnarray}{c}
%(\boldsymbol{\Phi}_{1}-\boldsymbol{\Phi}_{2})-(\boldsymbol{\Psi}_{1}-\boldsymbol{\Psi}_{2})=\boldsymbol{\Psi}_{1}\bigl(\mathbf{Q}_{2}-\mathbf{Q}_{1}\bigr)\bigl(\lambda_{2}\mathbf{I}-\mathbf{Z}_{2}\bigr)\nonumber\\ +\bigl(\lambda_{1}\mathbf{I}-\mathbf{Z}_{1}\bigr)\bigl(\mathbf{Q}_{2}-\mathbf{Q}_{1}\bigr)\boldsymbol{\Psi}_{2}+\bigl(\lambda_{1}\mathbf{I}-\mathbf{Z}_{1}\bigr)\bigl(\mathbf{Q}_{2}-\mathbf{Q}_{1}\bigr)\bigl(\lambda_{2}\mathbf{I}-\mathbf{Z}_{2}\bigr).\IEEEeqnarraynumspace\label{eq:Stp3_3} 
%\end{IEEEeqnarray}}
Substituting (\ref{eq:Stp3_3}) into (\ref{eq:StarredEq}) we can
rewrite the first term in the left-hand side of (\ref{eq:StarredEq})
as in (\ref{eq:13}) shown at the top of the following page. 

Comparing (\ref{eq:13}) and (\ref{eq:StarredEq}), our next step is to  bound the first two terms of the right-hand side (RHS) of (\ref{eq:13}) properly. To this end we multiply
both sides of (\ref{eq:KKT:stationary:reduced1}) with $\mathbf{Q}_{1}$,
and note that $\mathbf{Q}_{1}\mathbf{Z}_{1}=0$, which yields \setcounter{equation}{18}
\begin{equation}
\lambda_{1}\mathbf{Q}_{1}+\boldsymbol{\Psi}_{1}\mathbf{Q}_{1}=\boldsymbol{\Phi}_{1}\mathbf{Q}_{1}
\end{equation}
and thus we have{\thinmuskip=0mu 
\medmuskip =0mu
\thickmuskip =0mu
\begin{multline*}
\tr\bigl(\bigl(\lambda_{1}\mathbf{Q}_{1}+\boldsymbol{\Psi}_{1}\mathbf{Q}_{1}\bigr)\mathbf{Z}_{2}\mathbf{Q}_{1}\bigr)-\tr\bigl(\mathbf{Z}_{2}\mathbf{Q}_{1}\bigr)=\tr\bigl(\boldsymbol{\Phi}_{1}\mathbf{Q}_{1}\mathbf{Z}_{2}\mathbf{Q}_{1}\bigr)\\
-\tr(\mathbf{Z}_{2}\mathbf{Q}_{1})=\tr\bigl(\bigl(\boldsymbol{\Phi}_{1}\mathbf{Q}_{1}-\mathbf{I}\bigr)\mathbf{Z}_{2}\mathbf{Q}_{1}\bigr)=\tr\bigl(\bigl(\mathbf{I}+\boldsymbol{\Gamma}\bigr)^{-1}\boldsymbol{\Gamma}-\mathbf{I}\bigr)\mathbf{Z}_{2}\mathbf{Q}_{1}
\end{multline*}
}where $\boldsymbol{\Gamma}=\mathbf{H}\herm\mathbf{H}\mathbf{Q}_{1}$.
It is easy to see that $\bigl(\mathbf{I}+\boldsymbol{\Gamma}\bigr)^{-1}\boldsymbol{\Gamma}\preceq\mathbf{I}$
and thus $\bigl(\bigl(\mathbf{I}+\boldsymbol{\Gamma}\bigr)^{-1}\boldsymbol{\Gamma}-\mathbf{I}\bigr)\mathbf{Z}_{2}\mathbf{Q}_{1}\preceq\mathbf{0}$,
which leads to
\begin{equation}
\tr\bigl(\bigl(\lambda_{1}\mathbf{Q}_{1}+\boldsymbol{\Psi}_{1}\mathbf{Q}_{1}\bigr)\mathbf{Z}_{2}\mathbf{Q}_{1}\bigr)-\tr\bigl(\mathbf{Z}_{2}\mathbf{Q}_{1}\bigr)\leq0.\label{eq:Stp5_1}
\end{equation}
Similarly, we have
\begin{equation}
\tr\bigl(\bigl(\lambda_{2}\mathbf{Q}_{2}+\boldsymbol{\Psi}_{2}\mathbf{Q}_{2}\bigr)\mathbf{Z}_{1}\mathbf{Q}_{2}\bigr)-\tr\bigl(\mathbf{Z}_{1}\mathbf{Q}_{2}\bigr)\leq0.\label{eq:Stp5_2}
\end{equation}
Adding (\ref{eq:Stp5_1}) and (\ref{eq:Stp5_2}) gives 
\begin{align}
\tr\bigl((\lambda_{1}\mathbf{Q}_{1}+\boldsymbol{\Psi}_{1}\mathbf{Q}_{1})\mathbf{Z}_{2}\mathbf{Q}_{1}\bigr) & +\tr\bigl((\lambda_{2}\mathbf{Q}_{2}+\boldsymbol{\Psi}_{2}\mathbf{Q}_{2})\mathbf{Z}_{1}\mathbf{Q}_{2}\bigr)\nonumber \\
 & \hspace{-10pt}-\tr(\mathbf{Z}_{2}\mathbf{Q}_{1}+\mathbf{Z}_{1}\mathbf{Q}_{2})\leq0.\label{eq:21}
\end{align}
which results in (\ref{eq:14}) shown at the top of this page. It is not difficult to check that (\ref{eq:14}) holds due to (\ref{eq:21}) and also the fact that the last four terms of  the RHS of (\ref{eq:14}) are non-positive. Now subtracting both sides of (\ref{eq:13}) by $\tr(\mathbf{Z}_{2}\mathbf{Q}_{1}+\mathbf{Z}_{1}\mathbf{Q}_{2})$ and using (\ref{eq:14}) produces
\setcounter{equation}{23} {\small
%\thinmuskip=-1mu 
%\medmuskip =-1mu
%\thickmuskip =-1mu
\begin{IEEEeqnarray}{c}
\tr\bigl( \negmedspace \bigl(\negmedspace \bigl(\negmedspace\boldsymbol{\Phi}_{1}\negmedspace - \negmedspace \boldsymbol{\Phi}_{2}\!-\!\bigl(\boldsymbol{\Psi}_{1}-\boldsymbol{\Psi}_{2}\negmedspace\bigr)\negmedspace\bigr)\negmedspace\bigl(\negmedspace\mathbf{Q}_{1}-\mathbf{Q}_{2}\negmedspace\bigr)\negmedspace\bigr)\!-\!\tr\bigl(\negmedspace\mathbf{Z}_{2}\mathbf{Q}_{1}\!+\!\mathbf{Z}_{1}\mathbf{Q}_{2}\negmedspace\bigr)\negmedspace\leq \negmedspace 0.\IEEEeqnarraynumspace\label{eq:starred:contradict} 
\end{IEEEeqnarray}
}We remark that if $\mathbf{Q}_{1}\neq\mathbf{Q}_{2}$, then $ \lambda_{1}\lambda_{2}\tr\bigl((\mathbf{Q}_{2}-\mathbf{Q}_{1})(\mathbf{Q}_{2}-\mathbf{Q}_{1})\bigr) >0 $. Consequently, the inequality in (\ref{eq:14}) is
strict, and so is (\ref{eq:starred:contradict}), which contradicts (\ref{eq:StarredEq})
and completes the proof.
\end{IEEEproof}
An immediate consequence of Theorem \ref{thm:uniqueKKT} is that any
local optimization method is also a global optimization method for
the secrecy problem. In the next we exploit this property to derive
an efficient numerical method to solve (\ref{eq:secrecycapacity:org}).

\section{A Low-complexity Method}

\subsection{Algorithm Description}

%\begin{algorithm}[t]
%{
%%	\small
%\caption{Iterative algorithm for solving (\ref{eq:secrecycapacity:org})}
%\label{alg:AGPadaptive}
%\begin{algorithmic}[1]
%\REQUIRE $\mathbf{Y}_{1}=\mathbf{Q}_{0}\ensuremath{\in}\mathcal{\mathcal{\mathcal{Q}}}$,
%$\beta_{0}=L_{0}>0$, $\alpha=1$, $\xi\in(0,1)$, $\gamma_{u}>1$, $\epsilon>0$
%\REPEAT
%\REPEAT\label{alg:APMG:backtrack:start}
%\STATE $\mathbf{Q}_{k}=\Pi_{\mathcal{\mathcal{Q}}}\bigl(\mathbf{Y}_{k}+\frac{1}{\beta_{k-1}}\nabla C_{S}(\mathbf{Y}_{k})\bigr)$
%\label{alg:AGP:gradstep}
%\IF{ $C_{s}(\mathbf{Q}_{k})<\mu_{\beta_{k-1}}(\mathbf{Y}_{k};\mathbf{Q}_{k})$
%}
%\STATE $\beta_{k-1}=\gamma_{u}\beta_{k-1}$
%\ENDIF
%\UNTIL{ $C_{s}(\mathbf{Q}_{k})\geq\mu_{\beta_{k-1}}(\mathbf{Y}_{k};\mathbf{Q}_{k})$
%}\label{alg:APMG:backtrack:end}
%\STATE $\beta_{k}=\max\left\{ L_{0},\beta_{k-1}/\gamma_{u}\right\} $
%\STATE $\mathbf{Z}_{k}=\mathbf{Q}_{k}+\alpha(\mathbf{Q}_{k}-\mathbf{Q}_{k-1})$
%\label{alg:AGP:extrapolate}
%\IF{ $C_{s}(\mathbf{Z}_{k})\geq C_{s}(\mathbf{Q}_{k})$ and $\mathbf{Z}_{k}$
%is feasible}
%\STATE Update $\alpha=\min\left\{ \frac{\alpha}{\xi},1\right\} $
%and $\mathbf{Y}_{k+1}=\mathbf{Z}_{k}$\label{alg:AGP:goodextrapolation}
%\ELSE
%\STATE Update $\alpha=\xi\alpha$ and $\mathbf{Y}_{k+1}=\mathbf{Q}_{k}$\label{alg:AGP:badextrapolation}
%\ENDIF
%\UNTIL{ $|C_{s}(\mathbf{Q}_{k})|-|C_{s}(\mathbf{Q}_{k-1})|\leq\epsilon$
%}
%\ENSURE $\mathbf{Q}_{k}$
%\end{algorithmic}
%}
%\end{algorithm}

\begin{algorithm}[t]
{
		\caption{Iterative algorithm for solving (\ref{eq:secrecycapacity:org})}
	\label{alg:AGPadaptive}
\SetAlgoNoLine
\DontPrintSemicolon
\LinesNumbered 
\SetNlSty{textnormal}{}{}
\KwIn{ $\mathbf{Y}_{1}=\mathbf{Q}_{0}\ensuremath{\in}\mathcal{\mathcal{\mathcal{Q}}}$,
	$\beta_{0}=L_{0}>0$, $\alpha=1$, $\xi\in(0,1)$, $\gamma_{u}>1$, $\epsilon >0$, $k\leftarrow 1$}
\Repeat{ $|C_{s}(\mathbf{Q}_{k})|-|C_{s}(\mathbf{Q}_{k-1})|\leq\epsilon$}{
\Repeat(\tcc*[f]{line search}\label{alg:APMG:backtrack:start}){ $C_{s}(\mathbf{Q}_{k})\geq\mu_{\beta_{k-1}}(\mathbf{Y}_{k};\mathbf{Q}_{k})$\label{alg:APMG:backtrack:end}}{
	$\mathbf{Q}_{k}=\Pi_{\mathcal{\mathcal{Q}}}\bigl(\mathbf{Y}_{k}+\frac{1}{\beta_{k-1}}\nabla C_{S}(\mathbf{Y}_{k})\bigr)$ \label{alg:AGP:gradstep}\;
	\If{$C_{s}(\mathbf{Q}_{k})<\mu_{\beta_{k-1}}(\mathbf{Y}_{k};\mathbf{Q}_{k})$
	}{$\beta_{k-1}=\gamma_{u}\beta_{k-1}$\;}
}
$\beta_{k}=\max\left\{ L_{0},\beta_{k-1}/\gamma_{u}\right\} $\;
$\mathbf{Z}_{k}=\mathbf{Q}_{k}+\alpha(\mathbf{Q}_{k}-\mathbf{Q}_{k-1})$
\label{alg:AGP:extrapolate}\;
\eIf{$C_{s}(\mathbf{Z}_{k})\geq C_{s}(\mathbf{Q}_{k})$ and $\mathbf{Z}_{k}$
	is feasible}{Update $\alpha=\min\left\{ \frac{\alpha}{\xi},1\right\} $
	and $\mathbf{Y}_{k+1}=\mathbf{Z}_{k}$\label{alg:AGP:goodextrapolation}}
{
Update $\alpha=\xi\alpha$ and $\mathbf{Y}_{k+1}=\mathbf{Q}_{k}$\label{alg:AGP:badextrapolation}
}
$k\leftarrow k+1 $
}
\KwOut{$\mathbf{Q}_{k}$}	
}
\end{algorithm}

The proposed method is based on the accelerated projected gradient
method for non-convex programming with adaptive momentum
presented in \cite{li2017convergence}. The pseudo-code of the proposed
method is provided in Algorithm \ref{alg:AGPadaptive}, which is explained
in detail as follows. Let $\mathbf{Y}_{k}$ be the current operating
point. Then we take a projected gradient step to obtain the current
iterate $\mathbf{Q}_{k}$ (cf. Line \ref{alg:AGP:gradstep}). Note
that the notation $\Pi_{\mathcal{Q}}(\mathbf{X})$ in Line \ref{alg:AGP:gradstep}
denotes the projection of a given point $\mathbf{X}$ onto the feasible
set $\mathcal{Q}$, i.e., $\Pi_{\mathcal{Q}}(\mathbf{X})=\argmin \{ ||\mathbf{U}-\mathbf{X}||\ |\ \mathbf{U} \in \mathcal{Q} \} $. In contrast to \cite{li2017convergence} where
a constant  stepsize is used, we implement a backtracking line
search as done in Lines \ref{alg:APMG:backtrack:start}-\ref{alg:APMG:backtrack:end}
to find a proper step size, which is adopted from \cite{Nesterov2013}.
For this purpose we define a quadratic model of $C_{s}(\mathbf{Q})$
as
\begin{IEEEeqnarray}{rcl}
\mu_{\beta}(\mathbf{Q};\bar{\mathbf{Q}})&=&C_{s}(\mathbf{Q})\!+\!\tr\bigl(\nabla C_{s}(\mathbf{Q})\bigl(\bar{\mathbf{Q}}-\mathbf{Q}\bigr)\!\bigr)\!-\!\tfrac{\beta}{2}\bigl\Vert\bar{\mathbf{Q}}-\mathbf{Q}\bigr\Vert^{2}.\IEEEeqnarraynumspace
\end{IEEEeqnarray}
Recall that if $\beta\geq L$, where $L>0$ is a Lipschitz constant
of $\nabla C_{s}(\mathbf{Q})$ on $\mathcal{\mathcal{Q}}$, then the
inequality $C_{s}(\bar{\mathbf{Q}})\geq\mu_{\beta}(\mathbf{Q};\bar{\mathbf{Q}})$
holds \cite{Nesterov2013}. In the Appendix we show that $L=\sigma_{\max}^{2}\bigl(\mathbf{H}\herm\mathbf{H}\bigr)+\sigma_{\max}^{2}\bigl(\mathbf{G}\herm\mathbf{G}\bigr)$
is a Lipschitz constant for of $\nabla C_{s}(\mathbf{Q})$. To find
a proper $\beta$ in each iteration, we start from the value of $\beta$
in the previous iteration and increase it by $\gamma_{u}>1$ until
$\mu_{\beta}(\mathbf{Y}_{k};\mathbf{Q}_{k})$ becomes a lower bound
of $C_{s}(\mathbf{Q}_{k})$. In this way, the projected gradient step
always produces an improved iterate. Next, for acceleration, we compute
the extrapolated point $\mathbf{Z}_{k}=\mathbf{Q}_{k}+\alpha(\mathbf{Q}_{k}-\mathbf{Q}_{k-1})$,
where $\alpha$ is called the momentum parameter (cf. Line \ref{alg:AGP:extrapolate}).
For convex optimization, the momentum parameter is fixed. However,
since the objective in (\ref{eq:secrecycapacity:org}) is non-convex,
the extrapolation can be bad and thus $\alpha$ needs to be adapted
in accordance with the extrapolated point. To this end a monitor process
needs to be considered \cite{li2017convergence}. Specifically, if
the extrapolation reduces the current objective (i.e. bad extrapolation),
then the current iteration is taken for the next iteration and $\alpha$
is reduced with a rate $\xi$ (cf. Line \ref{alg:AGP:badextrapolation}).
Otherwise, the extrapolated point is taken to the next iteration and
$\alpha$ is increased by $1/\xi$ (cf. Line \ref{alg:AGP:goodextrapolation}). The stopping criterion for Algorithm \ref{alg:AGPadaptive} is when the increase in the last  iteration  is less than a small pre-determined parameter $ \epsilon $. 

 Algorithm \ref{alg:AGPadaptive} is very simple to implement because
$\nabla C_{s}(\mathbf{Q})$ is given in closed-form in (\ref{eq:gradF})
and the projection of a given point $\mathbf{X}$ onto $\mathcal{Q}$,
$\Pi_{\mathcal{Q}}(\mathbf{X})$, admits a water-filling like algorithm
as\cite{Ye2003}
\begin{equation}
\Pi_{\mathcal{Q}}(\mathbf{X})=\mathbf{U}\diag\bigl(\bigl[\mathbf{x}-c\bigr]_{+}\bigr)\mathbf{U}
\end{equation}
where $\mathbf{X}=\mathbf{U}\diag(\mathbf{x})\mathbf{U}\herm$ is
the eigenvalue decomposition of $\mathbf{X}$ and $c$ is the root
of the following equation 
\begin{equation}
\sum\nolimits _{i=1}^{N_{t}}\bigl[x_{i}-c\bigr]_{+}=P_{T}.
\end{equation}
%\begin{figure*}[t] 
%		\begin{minipage}{.32\textwidth}   
%			\centering   
%			\includegraphics[width = 0.98\linewidth, height = 4.1cm]{Figs/FigDegradedChan.pdf} 	
%			\caption{Convergence results of Algorithm~\ref{alg:AGPadaptive} for degraded channels.  $N_{t}=N_{r}=N_{e}=4$ at $P_{t}=15$ dB} 	
%			\label{fig:Fig1DegradedCh} 
%			\end{minipage} 
%	\hfill  
%		\begin{minipage}{.32\textwidth}   
%			\centering   
%			\includegraphics[width = 0.98\linewidth, height = 4.1cm]{Figs/FigNonDegradedChan.pdf} 	
%			\caption{Convergence results of Algorithm~\ref{alg:AGPadaptive} for non-degraded channels.  $N_{t}=N_{r}=N_{e}=4$ at $P_{t}=15$ dB} 	
%			\label{fig:Fig2NonDegradedCh} 
%		\end{minipage} 
%\hfill  
%		\begin{minipage}{.32\textwidth}   
%			\centering   
%			\includegraphics[width = 0.98\linewidth, height = 4.1cm]{Figs/Fig3Time.pdf} 
%			\caption{Time comparison of various algorithms for $N_t = 4$ and $N_r = 3$.\label{fig:FigTimeCompare}}
%			
%		\end{minipage} 
%\end{figure*}%
\begin{figure}
	\centering
	\subfloat[Degraded channels.]{\includegraphics[width = 0.85\columnwidth]{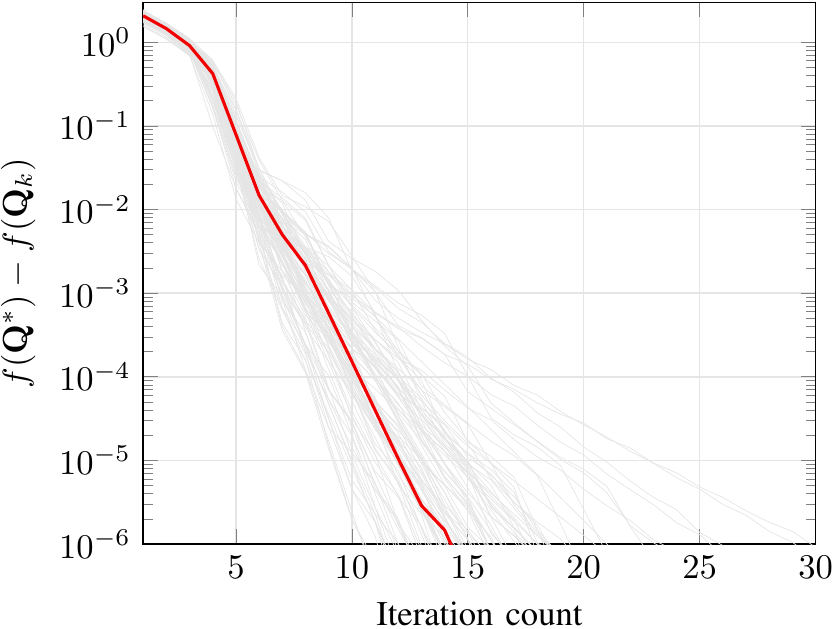}\label{fig:Fig1DegradedCh}  }
\vspace{0.5cm}
\subfloat[Non-degraded channels.]{\includegraphics[width = 0.85\columnwidth]{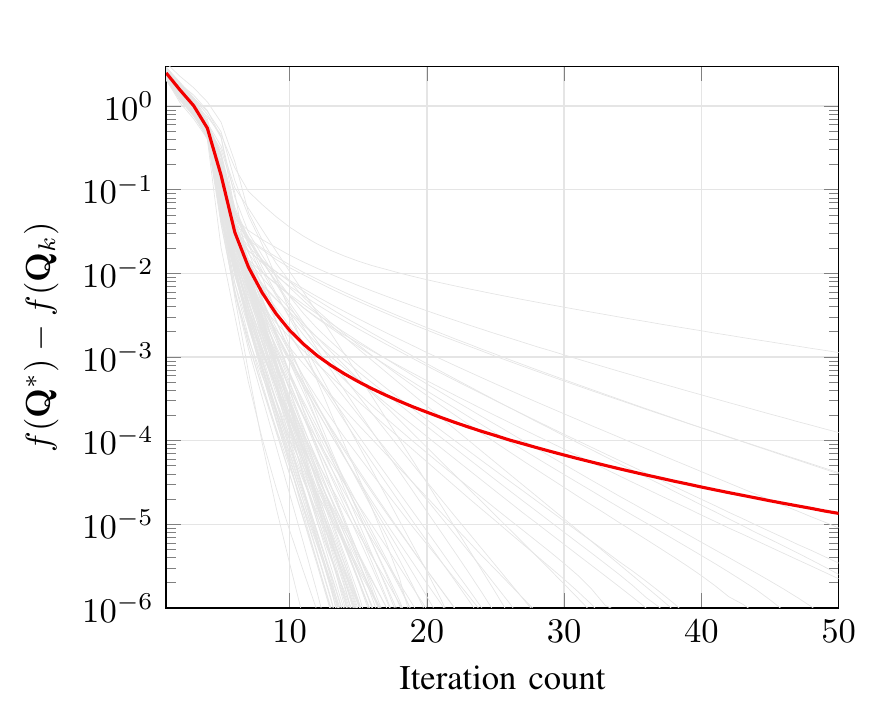} \label{fig:Fig2NonDegradedCh}}	

	\caption{Convergence results of Algorithm~\ref{alg:AGPadaptive}. $N_{t}=N_{r}=N_{e}=4$ at $P_{t}=15$ dB.}
	\label{fig:Convergence}
\end{figure}

\begin{figure}
	\centering 
	\includegraphics[width = 0.85\columnwidth]{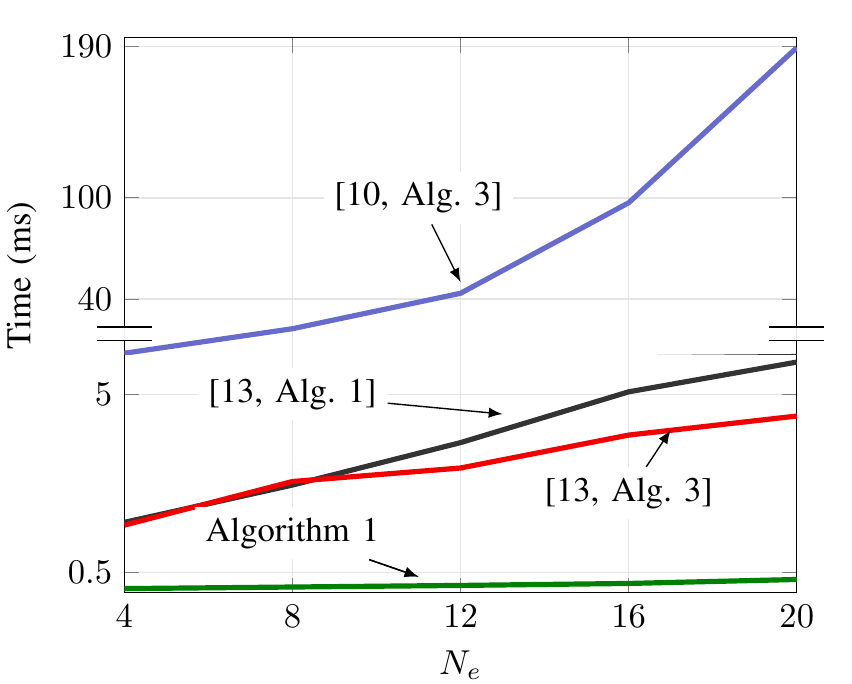} 
	\caption{Time comparison of various algorithms for $N_t = 4$ and $N_r = 3$.}
	\label{fig:FigTimeCompare}		
\end{figure}
\vspace{-0.1cm}
\subsection{Convergence Analysis}
We now show that the iterate sequence $\{\mathbf{Q}_{k}\}$ returned
by Algorithm \ref{alg:AGPadaptive} converges to a stationary of (\ref{eq:secrecycapacity:org}).
First, since $\nabla C_{s}(\mathbf{Q})$ is Lipschitz continuous,
the backtracking line search terminates in an finite number of steps.
More specifically, the number of line search steps at iteration $k\geq1$
is bounded by $2+\log_{\gamma_{u}}\bigl(\beta_{k}/\beta_{k-1}\bigr)$. Now
suppose that $C_{s}(\mathbf{Z}_{k})\geq C_{s}(\mathbf{Q}_{k})$ and
$\mathbf{Z}_{k}$ is feasible. Then we have $\mathbf{Y}_{k+1}=\mathbf{Z}_{k}$
(cf. Line \ref{alg:AGP:goodextrapolation}). The projection at
iteration $k+1$ can be explicitly written as 
%\begin{subequations}
\begin{IEEEeqnarray}{rcl}
\mathbf{Q}_{k+1} & = &\Pi_{\mathcal{\mathcal{Q}}}\bigl(\mathbf{Z}_{k}+\tfrac{1}{\beta_{k}}\nabla C_{s}(\mathbf{Z}_{k})\bigr) \IEEEyesnumber \IEEEyessubnumber\\
 & = &\underset{\mathbf{Q}\in\mathcal{\mathcal{Q}}}{\arg\min}\ \bigl\Vert\mathbf{Q}-\mathbf{Z}_{k}-\tfrac{1}{\beta_{k}}\nabla C_{s}(\mathbf{Z}_{k})\bigr\Vert^{2}\IEEEyessubnumber\\
 && \hspace{-2em}=\underset{\mathbf{Q}\in\mathcal{\mathcal{Q}}}{\arg\min}\ \tfrac{\beta_{k}}{2}\bigl\Vert\mathbf{Q}-\mathbf{Z}_{k}\bigr\Vert^{2}-\tr\bigl(\nabla C_{s}(\mathbf{Z}_{k})\bigl(\mathbf{Q}-\mathbf{Z}_{k}\bigr)\bigr).\IEEEyessubnumber\IEEEeqnarraynumspace\label{eq:project:k+1}
\end{IEEEeqnarray}
%\end{subequations}
Note that (\ref{eq:project:k+1}) implies 
\begin{equation}
\tfrac{\beta_{k}}{2}\bigl\Vert\mathbf{Q}_{k+1}-\mathbf{Z}_{k}\bigr\Vert^{2}-\tr\bigl(\nabla C_{s}(\mathbf{Z}_{k})\bigl(\mathbf{Q}_{k+1}-\mathbf{Z}_{k}\bigr)\bigr)\leq0\label{eq:optsolcond}.
\end{equation}
It follows from the condition in Line \ref{alg:APMG:backtrack:end}
that 
\begin{align}
 & C_{s}(\mathbf{Q}_{k+1})\geq\mu_{\beta_{k}}(\mathbf{Z}_{k};\mathbf{Q}_{k+1})=C_{s}(\mathbf{Z}_{k})\nonumber \\
 & +\tr\bigl(\nabla C_{s}(\mathbf{Z}_{k})\bigl(\mathbf{Q}_{k+1}-\mathbf{Z}_{k}\bigr)-\tfrac{\beta}{2}\bigl\Vert\mathbf{Q}_{k+1}-\mathbf{Z}_{k}\bigr\Vert^{2}
\end{align}
which, by using \eqref{eq:optsolcond}, yields
\begin{equation}
C_{s}(\mathbf{Q}_{k+1})\geq C_{s}(\mathbf{Z}_{k})\geq C_{s}(\mathbf{Q}_{k}).\label{eq:monotone}
\end{equation}
Similarly, if $\mathbf{Y}_{k+1}=\mathbf{Q}_{k}$, then we can also
prove that $C_{s}(\mathbf{Q}_{k+1})\geq C_{s}(\mathbf{Q}_{k})$ following
the same procedure. Note that the inequality is strict if $\mathbf{Q}_{k+1}\neq\mathbf{Q}_{k}$.
By noting that the feasible set is compact convex, we can conclude
the objective sequence $\{C_{s}(\mathbf{Q}_{k})\}$ is convergent
and there exists a subsequence $\{\mathbf{Q}_{k}\}$ converging to
a limit point $\mathbf{Q}^{\ast}$. The proof that $\mathbf{Q}^{\ast}$
is a stationary point of (\ref{eq:secrecycapacity:org}) is standard
and thus omitted here for the sake of brevity \cite{li2017convergence}.

\section{Numerical Results}

To illustrate Theorem \ref{thm:uniqueKKT} and also the convergence rate
of Algorithm~\ref{alg:AGPadaptive}, we plot the residual error (i.e.,
$C_{s}-C_{s}(\mathbf{Q}_{k})$) for both degraded and non-degraded
channels in Figs.~\ref{fig:Fig1DegradedCh} and~\ref{fig:Fig2NonDegradedCh}, respectively.
The channels $\mathbf{H}$ and $\text{\textbf{ G}}$ are generated
as $\mathcal{CN}(\mathbf{0,I})$. The secrecy capacity $C_{S}$ is
found using existing optimal algorithms. More specifically, for the
degraded MIMO WTC, problem (\ref{eq:secrecycapacity:org}) can be
reformulated as a standard semidefinite program and thus can be optimally
solved by off-the-shelf solvers such as MOSEK \cite{mosek}. For the
non-degraded case, we implement the barrier method \cite[Alg. 3]{lyoka2015minmaxEQ}
and \cite[Alg. 3]{Anshu:MIMOWTC:21VTC}, both of which can find the
secrecy capacity but are based on the equivalent convex-concave reformulation.
Note that Algorithm~\ref{alg:AGPadaptive} is applied to problem
(\ref{eq:secrecycapacity:org}) directly. As can be seen clearly in Fig. \ref{fig:Convergence},
Algorithm~\ref{alg:AGPadaptive} achieves monotonic convergence as
proved in (\ref{eq:monotone}). Also, the residual error is reduced
quickly to zero as the iteration process continues for both degraded
and non-degraded cases. These results indeed confirm that Algorithm~\ref{alg:AGPadaptive},
even when applied to the nonconvex form of the secrecy capacity problem,
can still compute the optimal solution, which is explained by Theorem
\ref{thm:uniqueKKT}.

To further demonstrate the benefit of Theorem \ref{thm:uniqueKKT}
and Algorithm~\ref{alg:AGPadaptive}, in Fig. \ref{fig:FigTimeCompare},
we compare the run time of Algorithm ~\ref{alg:AGPadaptive} as a
function of the number of the antennas at Eve. The simulation codes
are built on MATLAB and executed in a 64-bit Windows PC system with
16 GB RAM and Intel Core-i7, 3.20 GHz processor. 
%For this experiment,the numbers of antennas at Alice and Bob are $N_{t}=4$ and $N_{r}=3$.
We plotted the average actual run-time for 200 different channel realizations.
The stopping criteria for all the algorithms is when the increase
in the resulting objective is less than $10^{-5}$ during the last
5 iterations. We can see our proposed algorithm outperforms other
known methods such as \cite[Algorithm 1 and 3]{Anshu:MIMOWTC:21VTC} and
\cite[Algorithm 3]{lyoka2015minmaxEQ} in terms of time complexity.

\section{Conclusion}

We have proved that the secrecy rate maximization problem of the general MIMOME WTC (i.e. no assumption is made on whether the channel is degraded)
under a sum power constraint, despite its non-convexity, has a unique
KKT solution. The proof basically implies that any local optimization
method that aims to find a stationary solution can indeed solve secrecy
capacity problem for non-degraded MIMO wiretap channels which are known
to be non-convex. Motivated by this interesting result, we have also
presented an accelerated projected gradient method with adaptive momentum
to solve the secrecy problem. Simulation results have demonstrated
that the proposed algorithm can find the optimal solution very fast.

\appendix[Lipschitz constant of $\nabla C_{s}(\cdot)$ ]{Recall that $L>0$ is a Lipschitz constant of $\nabla C_{s}(\mathbf{Q})$
on $\mathcal{\mathcal{Q}}$ if the following inequality holds
\begin{equation}
\bigl\Vert\nabla C_{s}(\mathbf{X})-\nabla C_{s}(\mathbf{\mathbf{Y}})\bigr\Vert\leq L\bigl\Vert\mathbf{X}-\mathbf{Y}\bigr\Vert,\forall\mathbf{\mathbf{X}},\mathbf{Y}\in\mathcal{\mathcal{Q}}.
\end{equation}
Using (\ref{eq:gradF}) and the norm inequality, it is easy to see
that {\small
\begin{gather}
\bigl\Vert\nabla C_{s}(\mathbf{X})-\nabla C_{s}(\mathbf{Y})\bigr\Vert\leq\nonumber \\
\bigl\Vert\mathbf{H}\herm\bigl(\mathbf{I}+\mathbf{H}\mathbf{\mathbf{X}}\mathbf{H}\herm\bigr)^{-1}\mathbf{H}\bigl(\mathbf{Y}-\mathbf{X}\bigr)\mathbf{H}\herm\bigl(\mathbf{I}+\mathbf{H}\mathbf{Y}\mathbf{H}\herm\bigr)^{-1}\mathbf{H}\bigr\Vert+\nonumber \\
\bigl\Vert\mathbf{G}\herm\bigl(\mathbf{I}+\mathbf{G}\mathbf{X}\mathbf{G}\herm\bigr)^{-1}\mathbf{G}\bigl(\mathbf{Y}-\mathbf{X}\bigr)\mathbf{G}\herm\bigl(\mathbf{I}+\mathbf{G}\mathbf{Y}\mathbf{G}\herm\bigr)^{-1}\mathbf{G}\bigr\Vert.
\end{gather}
}We now recall the following well known inequality:
\begin{align}
||\mathbf{A}\mathbf{B}|| & \leq\lambda_{\max}(\mathbf{A})||\mathbf{B}||
\end{align}
where $\lambda_{\max}(\cdot)$ denotes the maximum singular value
of the matrix in the argument. Applying the above inequality and by
noting that $\lambda_{\max}\Bigl(\bigl(\mathbf{I}+\mathbf{H}\mathbf{\mathbf{X}}\mathbf{H}\herm\bigr)^{-1}\Bigr)\leq1$
we have
\begin{align*}
\bigl\Vert\nabla C_{s}(\mathbf{X})-\nabla C_{s}(\mathbf{Y})\bigr\Vert & \leq\bigl(\sigma\bigl(\mathbf{H}\herm\mathbf{H}\bigr)+\sigma_{\max}^{2}\bigl(\mathbf{G}\herm\mathbf{G}\bigr)\bigr)\bigl\Vert\mathbf{X}-\mathbf{Y}\bigr\Vert
\end{align*}
which means that $L=\sigma_{\max}^{2}\bigl(\mathbf{H}\herm\mathbf{H}\bigr)+\sigma_{\max}^{2}\bigl(\mathbf{G}\herm\mathbf{G}\bigr)$
is a Lipschitz constant of $\nabla C_{s}(\cdot)$.}
%\newpage
%\IEEEtriggeratref{4}
\bibliographystyle{IEEEtran}
\bibliography{IEEEabrv,paper}

\begin{thebibliography}{10}
\providecommand{\url}[1]{#1}
\csname url@samestyle\endcsname
\providecommand{\newblock}{\relax}
\providecommand{\bibinfo}[2]{#2}
\providecommand{\BIBentrySTDinterwordspacing}{\spaceskip=0pt\relax}
\providecommand{\BIBentryALTinterwordstretchfactor}{4}
\providecommand{\BIBentryALTinterwordspacing}{\spaceskip=\fontdimen2\font plus
\BIBentryALTinterwordstretchfactor\fontdimen3\font minus
  \fontdimen4\font\relax}
\providecommand{\BIBforeignlanguage}[2]{{%
\expandafter\ifx\csname l@#1\endcsname\relax
\typeout{** WARNING: IEEEtran.bst: No hyphenation pattern has been}%
\typeout{** loaded for the language `#1'. Using the pattern for}%
\typeout{** the default language instead.}%
\else
\language=\csname l@#1\endcsname
\fi
#2}}
\providecommand{\BIBdecl}{\relax}
\BIBdecl

\bibitem{Wyner75}
A.~D. Wyner, ``The wire-tap channel,'' \emph{Bell System Technical Journal},
  vol.~54, no.~8, pp. 1355--1387, 1975.

\bibitem{Gauss_wiretap}
S.~{Leung-Yan-Cheong} and M.~{Hellman}, ``The {Gaussian} wire-tap channel,''
  \emph{{IEEE} Trans. Inf. Theory}, vol.~24, no.~4, pp. 451--456, Jul. 1978.

\bibitem{ZangLi}
Z.~{Li}, W.~{Trappe}, and R.~{Yates}, ``Secret communication via multi-antenna
  transmission,'' in \emph{41st Annual Conference on Information Sciences and
  Systems 2007}, Mar. 2007, pp. 905--910.

\bibitem{Secrecy_cap_MISOME}
A.~{Khisti} and G.~W. {Wornell}, ``Secure transmission with multiple antennas
  {I}: The {MISOME} wiretap channel,'' \emph{{IEEE} Trans. Inf. Theory},
  vol.~56, no.~7, pp. 3088--3104, Jun. 2010.

\bibitem{MIMOME_WTC}
------, ``Secure transmission with multiple antennas part {II}: The {MIMOME}
  wiretap channel,'' \emph{{IEEE} Trans. Inf. Theory}, vol.~56, no.~11, pp.
  5515--5532, Oct. 2010.

\bibitem{Oggier2011SecCapEq}
F.~{Oggier} and B.~{Hassibi}, ``The secrecy capacity of the {MIMO} wiretap
  channel,'' \emph{{IEEE} Trans. Inf. Theory}, vol.~57, no.~8, pp. 4961--4972,
  Aug 2011.

\bibitem{Sec_MIMO_SPC_3}
S.~Fakoorian and A.~L. Swindlehurst, ``Full rank solutions for the {MIMO}
  {Gaussian} wiretap channel with an average power constraint,'' \emph{{IEEE}
  Trans. Signal Process.}, vol.~61, no.~10, pp. 2620--2631, Mar. 2013.

\bibitem{Li2013b}
Q.~Li, M.~Hong, H.-T. Wai, Y.-F. Liu, W.-K. Ma, and Z.-Q. Luo, ``Transmit
  solutions for {MIMO} wiretap channels using alternating optimization,''
  \emph{IEEE J. Sel. Areas Commun.}, vol.~31, no.~9, pp. 1714--1727, Sep. 2013.

\bibitem{MIMO_POTDC}
J.~Steinwandt, S.~A. Vorobyov, and M.~Haardt, ``{Secrecy rate maximization for
  MIMO Gaussian wiretap channels with multiple eavesdroppers via alternating
  matrix POTDC},'' in \emph{Proc. IEEE ICASSP 2014}, May 2014, pp. 5686--5690.

\bibitem{lyoka2015minmaxEQ}
S.~{Loyka} and C.~D. {Charalambous}, ``An algorithm for global maximization of
  secrecy rates in {Gaussian} {MIMO} wiretap channels,'' \emph{{IEEE} Trans.
  Commun.}, vol.~63, no.~6, pp. 2288--2299, Jun. 2015.

\bibitem{Loyka2016}
S.~Loyka and C.~D. Charalambous, ``{Optimal signaling for secure communications
  over Gaussian MIMO wiretap channels},'' in \emph{IEEE Trans. Inf. Theory},
  vol.~62, no.~12, Dec. 2016, pp. 7207--7215.

\bibitem{ThangNguyen2020}
T.~V. Nguyen, Q.-D. Vu, M.~Juntti, and L.-N. Tran, ``A low-complexity algorithm
  for achieving secrecy capacity in {MIMO} wiretap channels,'' in \emph{Proc.
  IEEE ICC 2020}, Jun. 2020.

\bibitem{Anshu:MIMOWTC:21VTC}
\BIBentryALTinterwordspacing
A.~Mukherjee, B.~Ottersten, and L.-N. Tran, ``Efficient numerical methods for
  secrecy capacity of {Gaussian MIMO} wiretap channel,'' in \emph{Proc. IEEE
  VTC 2021}, May 2021. [Online]. Available:
  \url{https://arxiv.org/abs/2102.10396}
\BIBentrySTDinterwordspacing

\bibitem{Anshu:MIMOWTC:2020}
\BIBentryALTinterwordspacing
------, ``On the {MIMO} secrecy capacity of {MIMO} wiretap channels: {C}onvex
  reformulation and efficient numerical methods,'' \emph{{submitted to IEEE
  Trans. Commun.}}, 2020. [Online]. Available:
  \url{https://arxiv.org/abs/2012.05667}
\BIBentrySTDinterwordspacing

\bibitem{Bernstein:2009}
D.~S. Bernstein, \emph{Matrix Mathematics: Theory, Facts, and Formulas (Second
  Edition)}.\hskip 1em plus 0.5em minus 0.4em\relax Princeton University Press,
  2009.

\bibitem{li2017convergence}
\BIBentryALTinterwordspacing
Q.~Li, Y.~Zhou, Y.~Liang, and P.~K. Varshney, ``Convergence analysis of
  proximal gradient with momentum for nonconvex optimization,'' in \emph{Proc.
  the 34th International Conference on Machine Learning}, vol.~70.\hskip 1em
  plus 0.5em minus 0.4em\relax PMLR, Aug. 2017, pp. 2111--2119. [Online].
  Available: \url{http://proceedings.mlr.press/v70/li17g.html}
\BIBentrySTDinterwordspacing

\bibitem{Nesterov2013}
Y.~Nesterov, ``Gradient methods for minimizing composite functions,''
  \emph{Math. Program.}, vol. 140, pp. 125--161, 2013.

\bibitem{Ye2003}
S.~Ye and R.~Blum, ``Optimized signaling for {MIMO} interference systems with
  feedback,'' \emph{{IEEE} Trans. Signal Process.}, vol.~51, no.~11, pp.
  2839--2848, Nov. 2003.

\bibitem{mosek}
\BIBentryALTinterwordspacing
M.~ApS, \emph{The MOSEK optimization toolbox for MATLAB manual. Version 9.2},
  2020. [Online]. Available:
  \url{https://docs.mosek.com/9.2/toolbox/index.html}
\BIBentrySTDinterwordspacing

\end{thebibliography}

\end{document}